\documentclass[11pt,a4paper]{article}
\usepackage[hyperref]{acl2021}
\usepackage{times}
\usepackage{latexsym}
\usepackage{graphicx}
\usepackage{subcaption}
\usepackage{booktabs}
\usepackage[export]{adjustbox}

\usepackage{enumitem}
\usepackage{multirow}

\usepackage{microtype}

\aclfinalcopy 
\def\aclpaperid{569} 

\title{MusicBERT: Symbolic Music Understanding with \\ Large-Scale Pre-Training}

\author{
    Mingliang Zeng,
    Xu Tan\thanks{$^*$Corresponding author: Xu Tan, xuta@microsoft.com}~, 
    Rui Wang,
    Zeqian Ju,
    Tao Qin,
    Tie-Yan Liu \\
    Microsoft Research Asia\\
    \texttt{\{v-minzeng,xuta,ruiwa,v-zeju,taoqin,tyliu\}@microsoft.com}
}

\date{}

\begin{document}
\maketitle
\begin{abstract}
Symbolic music understanding, which refers to the understanding of music from the symbolic data (e.g., MIDI format, but not audio), covers many music applications such as genre classification, emotion classification, and music pieces matching. While good music representations are beneficial for these applications, the lack of training data hinders representation learning. Inspired by the success of pre-training models in natural language processing, in this paper, we develop MusicBERT, a large-scale pre-trained model for music understanding. To this end, we construct a large-scale symbolic music corpus that contains more than 1 million music songs. Since symbolic music contains more structural (e.g., bar, position) and diverse information (e.g., tempo, instrument, and pitch), simply adopting the pre-training techniques from NLP to symbolic music only brings marginal gains. Therefore, we design several mechanisms, including OctupleMIDI encoding and bar-level masking strategy, to enhance pre-training with symbolic music data. Experiments demonstrate the advantages of MusicBERT on four music understanding tasks, including melody completion, accompaniment suggestion, genre classification, and style classification. Ablation studies also verify the effectiveness of our designs of OctupleMIDI encoding and bar-level masking strategy in MusicBERT.
\end{abstract}

\section{Introduction}

Music understanding, including tasks like genre classification, emotion classification, music pieces matching, has attracted lots of attention in both academia and industry. A better understanding of melody, rhythm, and music structure is not only beneficial for music information retrieval~\citep{casey2008content} but also helpful for music generation~\citep{huang2018music,sheng2020songmass}. Similar to natural language, music is usually represented in symbolic data format (e.g., MIDI)~\cite{jackendoff2009parallels,mcmullen2004music} with sequential tokens, and some methods~\cite{mikolov2013efficient,mikolov2013distributed} from NLP can be adopted for symbolic music understanding. Since the labeled training data for each music understanding task is usually scarce, previous works ~\citep{liang2020pirhdy,chuan2020context} leverage unlabeled music data to learn music token embeddings, similar to word embeddings in natural language tasks. Unfortunately, due to their shallow structures and limited unlabeled data, such embedding-based approaches have limited capability to learn powerful music representations.

In recent years, pre-trained language models (e.g., BERT) have been verified to be powerful for representation learning from large-scale unlabeled text corpora~\cite{devlin2018bert,radford2019language,yang2019xlnet,song2019mass,brown2020language,song2020mpnet}. However, it is challenging to directly apply the pre-training techniques from NLP to symbolic music because of the difference between natural text data and symbolic music data. First, since music songs are more structural (e.g., bar, position) and diverse (e.g., tempo, instrument, and pitch), encoding symbolic music is more complicated than natural language. The existing pianoroll-like~\cite{ji2020comprehensive} and MIDI-like~\cite{huang2020pop,ren2020popmag} representations of a song are too long to be processed by pre-trained models. For example, the length of a music song encoded by REMI~\cite{huang2020pop} has an average length of 15,679, as shown in Table~\ref{tab:avglen}. Due to the limits of computational resources, the length of sequences processed by a Transformer model is usually cropped to below 1,000. Thus such representations cannot capture sufficient information for song-level tasks. Accordingly, an effective, efficient, and universal symbolic music encoding method is needed for music representation learning. Second, due to the complicated encoding of symbolic music, the pre-training mechanism (e.g., the masking strategy like the masked language model in BERT) should be carefully designed to avoid information leakage in pre-training. Third, as pre-training relies on large-scale corpora, the lack of large-scale symbolic music corpora limits the potential of pre-training for music understanding.

\begin{table}[ht]
\small
    \centering
    \begin{tabular}{cccc}
        \toprule
        Encoding & OctupleMIDI & CP-like & REMI-like \\
        \midrule
        Tokens & \textbf{3607} & 6906 & 15679 \\
        \bottomrule
    \end{tabular}
    \caption{The average number of tokens per song on LMD dataset with different encoding methods.}
    \label{tab:avglen}
\end{table}

In this paper, we develop MusicBERT, a large-scale pre-trained model with carefully designed music encoding and masking strategy for music understanding.

\begin{itemize}[leftmargin=*]
    \item We design a novel music encoding method called OctupleMIDI, which encodes each note into a tuple with 8 elements. These 8 elements represent the different aspects of the characteristics of a musical note, including time signature, tempo, bar, position, instrument, pitch, duration, and velocity. OctupleMIDI has several advantages: 1) It largely reduces the length of a music sequence  (4x shorter than REMI~\cite{huang2020pop} and 2x shorter than CP~\cite{hsiao2021compound}), thus easing the modeling of music sequences by Transformer considering that music sequences themselves are very long. 2) It is note centric. Since each note contains the same 8-tuple structure and covers adequate information to express various music genres, such as changing time signature and long note duration, OctupleMIDI is much simpler and more universal than previous encoding methods.
    \item We carefully analyze the masking strategies for symbolic music understanding and propose a bar-level masking strategy for MusicBERT. The masking strategy in original BERT for NLP tasks randomly masks some tokens, which will cause information leakage in music pre-training. For example, some attributes are usually the same in a segment of consecutive tokens, such as time signature, tempo, instrument, bar, and position. Therefore, the masked tokens can be easily predicted by directly copying from the adjacent tokens since they are probably the same. Meanwhile, adjacent pitches usually follow the same chord so that a masked pitch token can be easily inferred from the adjacent tokens in the same chord. Therefore, we propose a bar-level masking strategy, which masks all the tokens of the same type (e.g., time signature, bar, instrument, or pitch) in a bar to avoid information leakage and encourage effective representation learning.
    \item Last but not least, we collect a large-scale and diverse symbolic music dataset, denoted as Million MIDI Dataset (MMD), that contains more than 1 million music songs, with different genres, including Rock, Electronic, Rap, Jazz, Latin, Classical, etc. To our knowledge, it is the largest in current literature, which is 10 times larger than the previous largest dataset LMD~\cite{raffel2016learning} in terms of the number of songs as shown in Table~\ref{tab:datasets}. Thus, this dataset greatly benefits representation learning for music understanding.
\end{itemize}

We fine-tune the pre-trained MusicBERT on four downstream music understanding tasks, including melody completion, accompaniment suggestion, genre classification, and style classification, and achieve state-of-the-art results on all the tasks. Furthermore, ablation studies show the effectiveness of the individual components in MusicBERT, including the OctupleMIDI encoding, the bar-level masking strategy, and the large-scale corpus.

The main contributions of this paper are summarized as follows:

\begin{itemize}[leftmargin=*]
    \item We pre-train MusicBERT on a large-scale symbolic music corpus that contains more than 1 million music songs and fine-tune MusicBERT on some music understanding tasks, achieving state-of-the-art results.
    \item We propose OctupleMIDI, an efficient and universal music encoding for music understanding, which leads to much shorter encoding sequences and is universal for various kinds of music.
    \item We design a bar-level masking strategy as the pre-training mechanism for MusicBERT, which significantly outperforms the naive token-level masking strategy used in natural language pre-training.
\end{itemize}

\section{Related Works}

\subsection{Symbolic Music Understanding}

Inspired by word2vec~\cite{mikolov2013efficient,mikolov2013distributed} in NLP, previous works on symbolic music understanding learn music embedding by predicting a music symbol based on its neighborhood symbols. \citet{huang2016chordripple,madjiheurem2016chord2vec} regard chords as words in NLP and learn chords representations using the word2vec model. \citet{herremans2017modeling,chuan2020context,liang2020pirhdy} divide music pieces into non-overlapping music slices with a fixed duration and train the embeddings for each slice. ~\citet{hirai2019melody2vec} cluster musical notes into groups and regard such groups as words for representation learning. However, the word2vec-based approaches mentioned above only use relatively small neural network models and take only a few (usually 4-5) surrounding music tokens as inputs, which have limited capability compared with recently developed deep and big pre-trained models like BERT~\cite{devlin2018bert}, which takes a long sentence (e.g., with 512 words/tokens) as input. In this paper, we pre-train big/deep models over a large-scale music corpus and use more context as input to improve symbolic music understanding.

\subsection{Symbolic Music Encoding}
There are two main approaches to encode symbolic music: pianoroll-based and MIDI-based.

In pianoroll-based methods~\cite{ji2020comprehensive,brunner2018symbolic}, music is usually encoded into a 2-dimensional binary matrix, where one dimension represents pitches, and the other represents time steps. Each element in the matrix indicates whether the pitch is played at that time step. As a result, a note is always divided into multiple fixed intervals, which is inefficient, especially for long notes.

MIDI is a technical standard for transferring digital instrument data. Many works in symbolic music~\cite{oore2020time,huang2018music} encode music pieces based on MIDI events, including note-on, note-off, time-shift, etc. REMI~\cite{huang2020pop} improves the basic MIDI-like encoding using note-duration, bar, position, chord, and tempo. Inspired by REMI~\cite{huang2020pop}, PopMAG~\cite{ren2020popmag} and Compound Word (CP)~\cite{hsiao2021compound} compress the attributes of a note, including pitch, duration, and velocity, into one symbol and reduces duplicated position events. Although such MIDI-like approaches avoid redundancy for long notes, they still need multiple tokens to represent the attributes, position, and metadata of a single note, which can be further compressed. This paper proposes OctupleMIDI, a MIDI-based encoding method, which is efficient due to the reduced sequence length and universal to support various music genres.

\subsection{Masking Strategies in Pre-training}

Masking strategies play a key role in NLP pre-training. For example, BERT~\cite{devlin2018bert} and RoBERTa ~\cite{liu2019roberta} randomly mask some tokens in an input sequence and learn to predict the masked tokens. Furthermore, since adjacent tokens may form a word or a phrase, some works consider masking consecutive tokens. For example, MASS~\cite{song2019mass} randomly masks a fragment of several consecutive tokens in the input, and SpanBERT~\cite{joshi2020spanbert} randomly masks contiguous spans instead of tokens. However, symbolic music is different from language. First, symbolic music contains structural (e.g., bar, position) and diverse information (e.g., tempo, instrument, and pitch), while natural language can be regarded as homogeneous data, which only contains text. Second, music and language follow different rules. Specifically, the language rules include grammar and spelling, while the music rules include beat, chord, etc. Accordingly, the masking strategies for symbolic music need to be specifically designed; otherwise, it may limit the potential of pre-training because of information leakage, as we analyzed before. In this paper, we carefully design a bar-level masking strategy for symbolic music pre-training.

\section{Methodology}

In this section, we introduce MusicBERT, a large-scale Transformer model for symbolic music understanding. We first overview the model structure and then describe the OctupleMIDI encoding and masking strategy for pre-training. At last, we describe the large-scale music corpus with over 1 million songs used in MusicBERT pre-training.

\begin{figure*}[ht]
    \centering
    \includegraphics[page=4,width=0.8\textwidth,trim=60 80 60 125, clip]{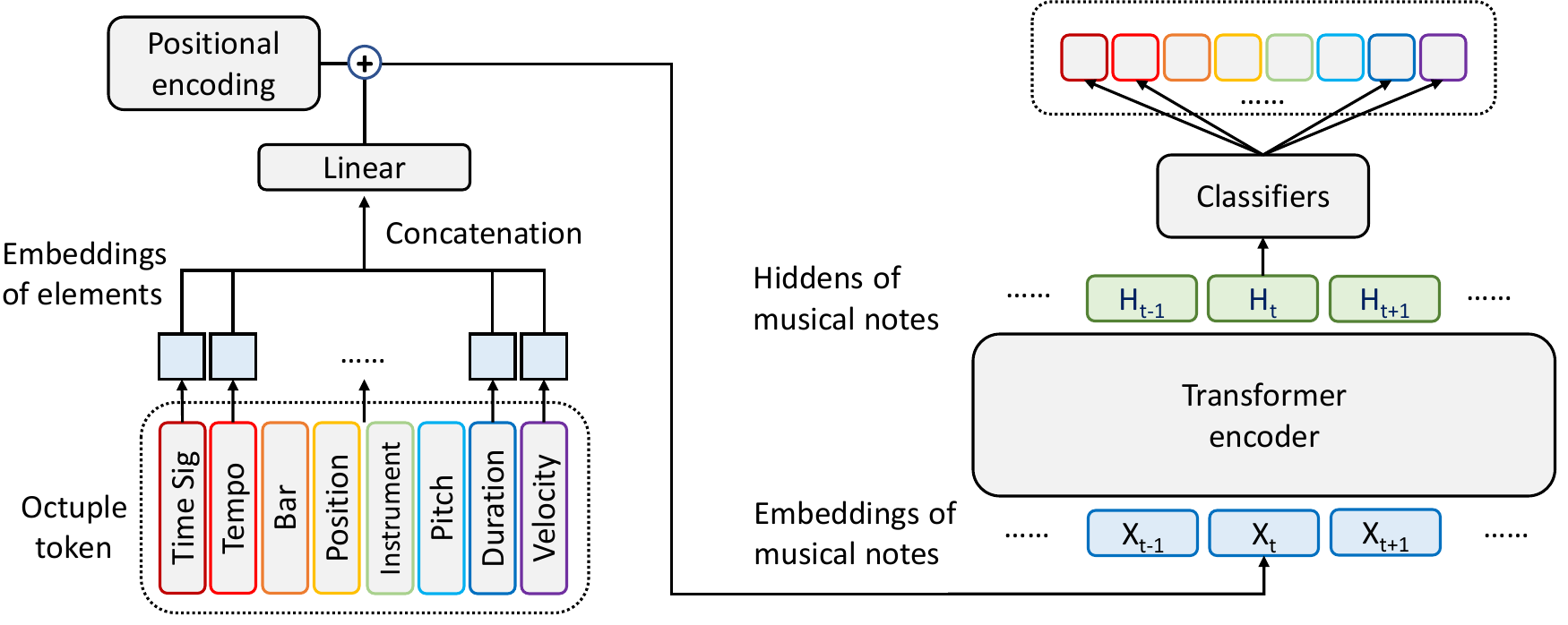}
    \caption{Model structure of MusicBERT.}
    \label{fig:cpmodel}
\end{figure*}

\subsection{Model Overview}

As shown in Figure~\ref{fig:cpmodel}, MusicBERT pre-trains a Transformer encoder~\cite{vaswani2017attention, devlin2018bert}, with masked language modeling where some tokens in the input music sequence are masked and are predicted in the model output. To encode the music sequence more efficiently, we propose a novel encoding method called OctupleMIDI, which encodes a symbolic music piece into a sequence of octuple tokens (an 8-tuple) that contains 8 basic elements related to a music note (we introduce OctupleMIDI in detail in Sec. \ref{sec:compoundToken}). To convert the octuple tokens in each sequence step into the input of the Transformer encoder, we concatenate the embeddings of the 8 elements and use a linear layer to convert them into a single vector. Then, the converted vector is added with the corresponding position embeddings and taken as the input of the Transformer encoder. To predict each of the 8 tokens in the 8-tuple from the Transformer encoder, we add 8 different softmax layers to map the hidden of the Transformer encoder to the vocabulary sizes of 8 different element types, respectively.

\subsection{OctupleMIDI Encoding}
\label{sec:compoundToken}

\begin{figure*}[ht!]
    \centering
    \begin{subfigure}[b]{0.49\textwidth}
        \includegraphics[page=1,width=1\textwidth,trim=195 80 200 100, clip]{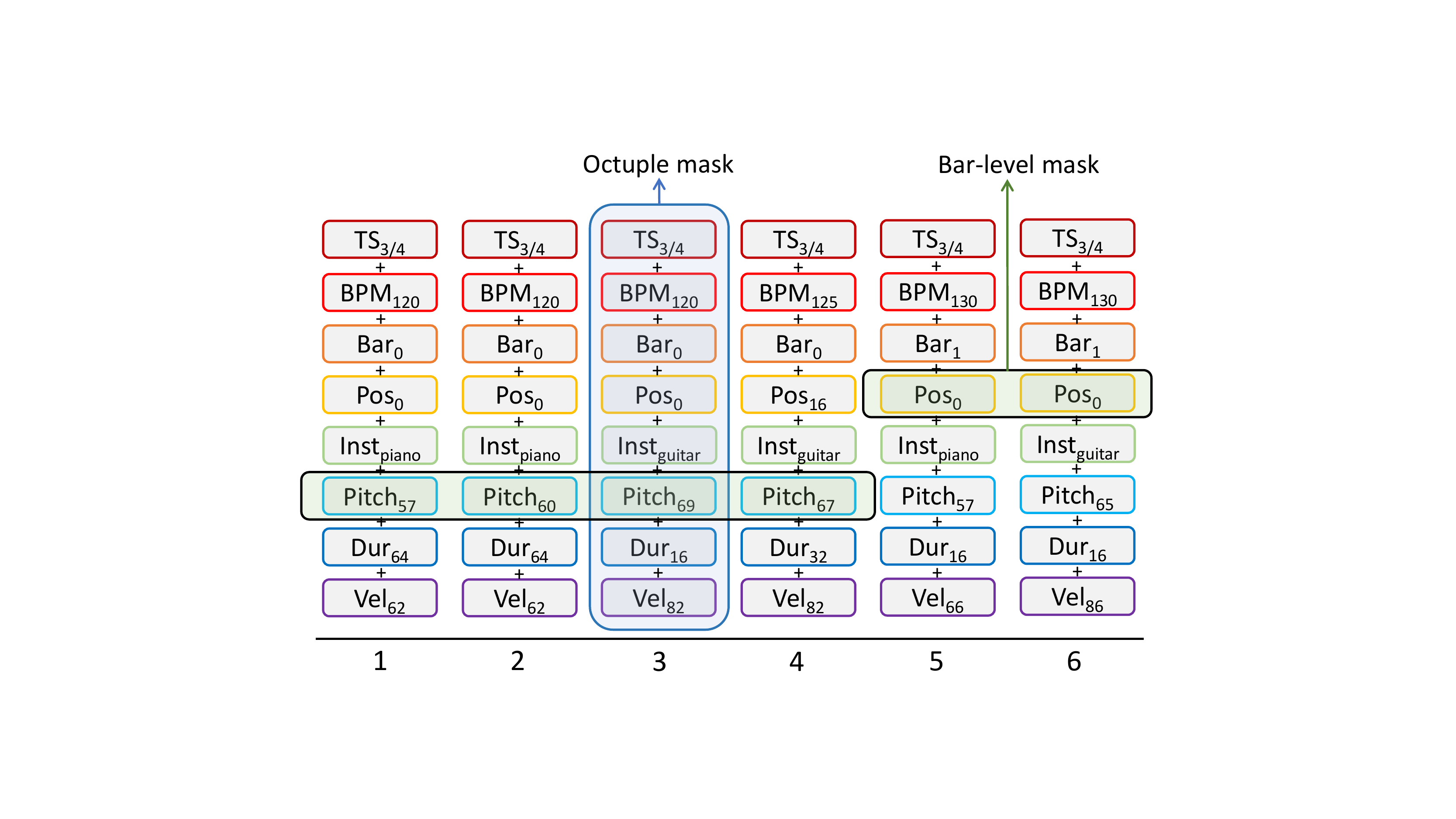}
        \caption{OctupleMIDI encoding.}
        \label{fig:OctupleMIDI}
    \end{subfigure}
    \begin{subfigure}[b]{0.49\textwidth}
        \includegraphics[page=2,width=1\textwidth,trim=210 125 265 125, clip]{MusicBERTFigure.pdf}
        \caption{CP-Like encoding.}
        \label{fig:REMI}
    \end{subfigure}
    \begin{subfigure}{\textwidth}
        \centering
        \includegraphics[page=3,width=\textwidth,trim=0 210 0 205, clip]{MusicBERTFigure.pdf}
        \caption{REMI-Like encoding.}
        \label{fig:MusicBERT}
    \end{subfigure}
    \caption{Different encoding methods for symbolic music. }
    \label{fig:encoding}
\end{figure*}

Previous works ~\cite{liang2020pirhdy} encode the symbolic music in a pianoroll-like way, which is not efficient since a note is always divided into multiple fixed small intervals (e.g., a quarter note is represented with 8 consecutive tokens). MIDI-like approaches~\cite{huang2020pop,ren2020popmag,hsiao2021compound} encode a note into several tokens based on MIDI events, making encoding much shorter, and has been widely used in music generation tasks. However, previous MIDI-like representations are still long for the Transformer structure due to computation complexity and learning efficiency. Accordingly, we propose a compact symbolic music encoding method called OctupleMIDI for music understanding tasks. As shown in Fig. \ref{fig:encoding}, OctupleMIDI encodes 6 notes into 6 tokens, which is much shorter than 33 tokens with REMI~\cite{huang2020pop} and 16 tokens with CP~\cite{hsiao2021compound}. Meanwhile, OctupleMIDI is general for various kinds of music. For example, OctupleMIDI supports changeable time signature and tempo.

In OctupleMIDI, we use sequences of octuple tokens to represent symbolic music. Each octuple token corresponds to a note and contains 8 elements, including time signature, tempo, bar, position, instrument, pitch, duration, and velocity. We introduce the details of each element as follows:

\begin{itemize}[leftmargin=*]
    \item Time signature. A time signature is denoted as a fraction (e.g., 2/4), where the denominator is a power of two in range [1, 64], representing the length of a beat (measured by note duration, e.g., a quarter note in 2/4), and the numerator is an integer in range [1, 128], representing the number of beats in a bar (e.g., 2 beats in 2/4). The value of the fraction measures the duration of a bar normalized by a whole note (e.g., 2/4 means that the duration of a bar is a half note). We consider the duration of a bar is no more than two whole notes. Otherwise, we divide a long bar into several equal-duration bars no longer than two whole notes. Therefore, there are 254 different valid time signatures in OctupleMIDI.
    \item Tempo. Tempo is measured in beats per minute (BPM), which describes the pace of music. In most music samples, tempo values are in range [24, 200]. For OctupleMIDI encoding, we quantize tempo values to 49 different values from 16 to 256, forming a geometric sequence.
    \item Bar and position. We use bar and position to indicate the on-set time of a note hierarchically. In the coarse level, we use 256 tokens ranging from 0 to 255 to represent the bar, supporting up to 256 bars in a music piece, which is sufficient in most cases. In the fine-grained level (inside each bar), we use position with a granularity of 1/64 note to represent the on-set time of a note, starting from 0 in each bar. Therefore, we need 128 tokens to represent position since the duration of a bar is no more than two whole notes, as described above. For example, in a bar with a time signature of 3/4, the possible value of position is from 0 to 47.
    \item Instrument. According to the MIDI format, we use 129 tokens to represent instruments, where 0 to 127 stands for different general instruments such as piano and bass, and 128 stands for the special percussion instrument such as drum.
    \item Pitch. For notes of general instruments, we use 128 tokens to represent pitch values following the MIDI format. However, for notes of percussion instruments, there are no pitches but percussion types (e.g., bass drum, hand clap). Therefore, we use another 128 ``pitch" tokens to represent percussion type for percussion instruments.
    \item Duration. To support the long note duration (up to 60 whole notes in the common music genre) with a fixed set of duration tokens, we propose a mixed resolution method: using high resolution (e.g., sixty-fourth note) when the note duration is small and using a low resolution (e.g., thirty-second note or larger) when the note duration is large. Specifically, we use 128 tokens to represent duration, starting from 0, with an increment of sixty-fourth note for the first 16 tokens, and double the increment (i.e., thirty-second note) every time for next 16 tokens. The duration for percussion instruments is meaningless, so we always set them to 0. 
    \item Velocity. We quantize the velocity of a note in the MIDI format into 32 different values with an interval of 4 (i.e., 2, 6, 10, 14, \ldots, 122, 126).
\end{itemize}

An example of a music sequence in OctupleMIDI encoding is shown in Fig. \ref{fig:OctupleMIDI}.













\subsection{Masking Strategy}
\label{sec_mask_strategy}

Inspired by the masked language model in BERT~\cite{devlin2018bert}, we randomly mask some elements in the input sequence of octuple tokens and predict the masked ones. A naive masking strategy is to randomly mask some octuple tokens (mask all the elements in an octuple token), which is denoted as octuple masking as shown in Fig. \ref{fig:OctupleMIDI}. However, considering the specific rules of music, such naive strategy will cause information leakage, and thus cannot well learn the contextual representation. 

A music song consists of multiple bars, which can be regarded as highly internally related units. An octuple token can be easily inferred from the adjacent tokens in the same bar. Specifically, time signature, tempo, and bar usually remain the same in the same bar. Instrument and position values in the same bar follow regular patterns, where the instrument is limited to a small-scale fixed set of values, and the position values are non decreasing. Moreover, a chord is a fixed combination of pitches, which always appear in adjacent positions. Accordingly, we propose a novel bar-level masking strategy, where the elements with the same type in the same bar are regarded as a unit and are masked simultaneously. In this way, information leakage can be avoided, and better contextual representation can be learned through pre-training. An example of bar-level masking is shown in Fig. \ref{fig:OctupleMIDI}.

For the masked elements, 80\% of them are replaced with $[MASK]$, 10\% of them are replaced with a random element, and 10\% remain unchanged, following the common practice~\cite{devlin2018bert,joshi2020spanbert,liu2019roberta}. Inspired by RoBERTa~\cite{liu2019roberta}, we remove the next sentence prediction task in pre-training and adopt a dynamic masking strategy, where the masked sequence is generated every time when feeding a sequence to the model.

\subsection{Pre-training Corpus}

A large-scale music dataset is necessary to learn good music representations from pre-training. However, previous symbolic music datasets are usually of small scale: 1) the MAESTRO dataset ~\cite{hawthorne2018enabling} contains only one thousand piano performances; 2) the GiantMIDI-Piano dataset ~\cite{kong2020giantmidi,kong2020high} contains slightly larger but still only ten thousands of piano performances; 3) the largest open-sourced symbolic music dataset by now is the Lakh-MIDI Dataset (LMD)~\cite{raffel2016learning}, which contains about 100K songs. 

To pre-train a powerful model with good music representations, we build a large-scale symbolic music corpus with over 1.5 million songs. Specifically, we first crawled a large amount of music files, cleaned files that are malformed or blank, and then converted those files into our symbolic music encoding. Since OctupleMIDI encoding is universal, most MIDI files can be converted to our encoding without noticeable loss of musical information. We found that these music files may have almost identical music content even if their hash values are different. Therefore, we developed an efficient way to deduplicate them: we first omitted all elements except instrument and pitch in the encoding, then got hash values of the remaining sequence and use it as the fingerprint of this music file, which is further used for deduplication. After cleaning and deduplication, we obtained  1.5 million songs with 2 billion octuple tokens (musical notes). We denote our dataset as Million-MIDI Dataset (MMD). We compare the sizes of different music datasets in Table~\ref{tab:datasets}.

\begin{table}[ht]
\small
    \centering
    \begin{tabular}{llll}
        \toprule
        Dataset & Songs & Notes (Millions) \\
        \midrule
        MAESTRO & 1,184 & 6 \\
        GiantMIDI-Piano & 10,854 & 39 \\
        LMD & 148,403 & 535 \\
        \midrule
        MMD & \textbf{1,524,557} & \textbf{2,075} \\
        \bottomrule
    \end{tabular}
    \caption{The sizes of different music datasets. Since LMD also consists of MIDI files from various websites, we perform the same cleaning and deduplication process as used in MMD and get 148,403 songs in LMD.
}
    \label{tab:datasets}
\end{table}

\vspace{-0.2cm}

\section{Experiments and Results}

In this section, we first introduce the pre-training setup for MusicBERT, and then fine-tune MusicBERT on several downstream music understanding tasks to compare it with previous approaches. Finally, more method analyses are conducted to verify the effectiveness of the designs in MusicBERT. 

\begin{table*}[ht]
\small
    \centering
    \begin{tabular}{lcccccccccccc}
        \toprule
         \multirow{3}{*}{Model} & \multicolumn{5}{c}{Melody Completion} & \multicolumn{5}{c}{Accompaniment Suggestion} & \multicolumn{2}{c}{Classification} \\
         \cmidrule(lr){2-6} \cmidrule(lr){7-11} \cmidrule(lr){12-13}
         & \multirow{2}{*}{MAP} & HITS & HITS & HITS & HITS & \multirow{2}{*}{MAP} & HITS & HITS & HITS & HITS & Genre & Style \\
         & & @1 & @5 & @10 & @25 & & @1 & @5 & @20 & @25 & F1 & F1 \\
        \midrule
         \textbf{melody2vec\textsubscript{F}} & 0.646 & 0.578 & 0.717 & 0.774 & 0.867 & - & - & - & - & - & 0.649 & 0.299 \\
         \textbf{melody2vec\textsubscript{B}} & 0.641 & 0.571 & 0.712 & 0.772 & 0.866 & - & - & - & - & - & 0.647 & 0.293\\
         \textbf{tonnetz} & 0.683 & 0.545 & 0.865 & 0.946 & 0.993 & 0.423 & 0.101 & 0.407 & 0.628 & 0.897 & 0.627 & 0.253\\
         \textbf{pianoroll} & 0.762 & 0.645 & 0.916 & 0.967 & 0.995 & 0.567 & 0.166 & 0.541 & 0.720 & 0.921 & 0.640 & 0.365 \\
         \textbf{PiRhDy\textsubscript{GH}} & 0.858 & 0.775 & 0.966 & 0.988 & 0.999 & 0.651 & 0.211 & 0.625 & 0.812 & 0.965 & 0.663 & 0.448\\
         \textbf{PiRhDy\textsubscript{GM}} & 0.971 & 0.950 & 0.995 & 0.998 & 0.999 & 0.567 & 0.184 & 0.540 & 0.718 & 0.919 & 0.668 & 0.471 \\
         \midrule
         \textbf{MusicBERT\textsubscript{small}} & 0.982 & 0.971 & 0.996 & 0.999 & 1.000 & 0.930 & 0.329 & 0.843 & 0.993 & 0.997 & 0.761 & 0.626 \\
         \textbf{MusicBERT\textsubscript{base}} & \textbf{0.985} & \textbf{0.975} & \textbf{0.997} & \textbf{0.999} & \textbf{1.000} & \textbf{0.946} & \textbf{0.333} & \textbf{0.857} & \textbf{0.996} & \textbf{0.998} & \textbf{0.784} & \textbf{0.645} \\
        \bottomrule
    \end{tabular}
    \caption{Results of different models on the four downstream tasks: melody completion, accompaniment suggestion, genre classification, and style classification. We choose four baseline models: Melody2vec~\cite{hirai2019melody2vec} is widely used in music understanding tasks, tonnetz~\cite{chuan2018modeling} and pianoroll~\cite{dong2018pypianoroll} are classical methods for music representation, PiRhDy~\cite{liang2020pirhdy} is a new model that significantly outperforms previous models in all four downstream tasks. Results of melody2vec on accompaniment suggestion task are emitted since it only encodes melody part of music.}
    \label{tab:scores}
\end{table*}

\subsection{Pre-training Setup}
\paragraph{Model Configuration} We pre-train two versions of MusicBERT: 1) MusicBERT\textsubscript{small} on the small-scale LMD dataset, which is mainly for a fair comparison with previous works on music understanding such as PiRhDy~\cite{liang2020pirhdy} and melody2vec~\cite{hirai2019melody2vec}, which are also pre-trained on LMD; 2) MusicBERT\textsubscript{base} on the large-scale MMD dataset, for pushing the SOTA results and showing the scalability of MusicBERT. The details of the two MusicBERT models are shown in Table~\ref{tab:modelsize}. We use our proposed bar-level masking strategy with a masking probability of 15\%. In addition, we use tokens of 8 duplicated elements to represent the class token and the end of sequence token. They are also masked with a 15\% probability. 

\begin{table}[ht]
\small
    \centering
    \begin{tabular}{lll}
        \toprule
        MusicBERT & small & base \\
        \midrule
        Number of layers & 4 & 12 \\
        Element embedding size & 512 & 768 \\
        Hidden size & 512 & 768 \\
        FFN inner hidden size & 2048 & 3072 \\
        \#Attention heads & 8 & 12 \\
        Pre-training dataset & LMD & MMD \\
        \bottomrule
    \end{tabular}
    \caption{The model configurations of MusicBERT.}
    \label{tab:modelsize}
\end{table}

\paragraph{Pre-training Details} The average sequence length of OctupleMIDI representation of a song is 3607 tokens as shown in Table~\ref{tab:avglen}, which is too long to model in Transformer. Therefore, we randomly sample segments with a length of 1024 tokens for pre-training. Following \citet{liu2019roberta}, we pre-train MusicBERT on 8 NVIDIA V100 GPUs for 4 days, and there are 125,000 steps in total, with a batch size of 256 sequences, each has a maximum length of 1024 tokens. We use Adam~\cite{kingma2014adam} optimizer with $\beta _1$=0.9, $\beta _2$=0.98, $\epsilon$=1e-6 and $L_2$ weight decay of 0.01. The learning rate is warmed up over the first 25,000 steps to a peak value of 5e-4 and then linearly decayed. Dropout value on all layers and attention weights are set to 0.1. 

\subsection{Fine-tuning MusicBERT}
We fine-tune MusicBERT on four downstream tasks: two phrase-level tasks (i.e., melody completion and accompaniment suggestion) and two song-level tasks (i.e., genre and style classification). For the two phrase-level tasks, the learning rate is warmed up over the first 50,000 steps to a peak value of 5e-5 and then linearly decayed until reaching 250,000 total updates. For the two song-level tasks, the learning rate is warmed up over the first 4,000 steps to a peak value of 5e-5 and then linearly decayed until reaching 20,000 total updates. The batch size is set to 64 sequences for both tasks. Other settings are the same as pre-training.
We compare MusicBERT with previous works on symbolic music understanding, including PiRhDy~\cite{liang2020pirhdy} and melody2vec~\cite{hirai2019melody2vec}.

\subsubsection{Melody Completion}
Melody completion~\cite{liang2020pirhdy} is to find the most matched consecutive phrase in a given set of candidates for a given melodic phrase. There are 1,793,760 data pairs in the training set and 198,665 data groups in the test set in this task~\cite{liang2020pirhdy}. Each training data pair consists of one positive sample and one negative sample, while each test data group consists of 1 positive sample and 49 negative samples. We use mean average precision (MAP) and HITS@k (k=1, 5, 10, 25, indicating the rate of correctly chosen phrase in the top k candidates) as evaluation metrics, making comparisons with PiRhDy~\cite{liang2020pirhdy}, melody2vec~\cite{hirai2019melody2vec}, pianoroll~\cite{dong2018pypianoroll}, tonnetz~\cite{chuan2018modeling}. As shown in Table~\ref{tab:scores}, MusicBERT\textsubscript{small} outperforms all previous works on the same pre-training dataset LMD, indicating the advantage of MusicBERT on learning representations from melodic context. MusicBERT\textsubscript{base} with a larger model and pre-training corpus can further achieve better results, showing the effectiveness of large-scale pre-training.

\subsubsection{Accompaniment Suggestion}
Accompaniment suggestion~\cite{liang2020pirhdy} is to find the most related accompaniment phrase in a given set of harmonic phrase candidates for a given melodic phrase. There are 7,900,585 data pairs in the training set, each consisting of one positive sample and one negative sample, and 202,816 data groups in the test set~\cite{liang2020pirhdy}. Each group in the test set consists of N positive samples and (50-N) negative samples when there are N accompaniment tracks in the MIDI file of that sample. We use mean average precision (MAP) and HITS@k as metrics, making comparisons with PiRhDy~\cite{liang2020pirhdy}, pianoroll~\cite{dong2018pypianoroll}, tonnetz~\cite{chuan2018modeling}. As shown in Table~\ref{tab:scores}, MusicBERT models perform much better than previous works, indicating the advantages of MusicBERT in understanding harmonic context.

\subsubsection{Genre and Style Classification}

Genre classification and style classification~\citep{ferraro2018large} are multi-label classification tasks. Following ~\citet{ferraro2018large}, we use the TOP-MAGD dataset for genre classification and the MASD dataset for style classification. TOP-MAGD contains 22,535 annotated files of 13 genres, and MASD contains 17,785 files of 25 styles. We evaluate MusicBERT on TOP-MAGD and MASD using 5-fold cross-validation and use the F1-micro score as the metric~\cite{liang2020pirhdy,ferraro2018large,oramas2017multi}. Due to the limitation of computational resources, for songs with more than 1,000 octuple tokens, we randomly crop segments with 1,000 tokens. On average, the selected segment covers more than 1/4 of a music song according to Table~\ref{tab:avglen}, which is enough to capture sufficient information for identifying genres and styles. As shown in Table~\ref{tab:scores}, MusicBERT models significantly outperform previous works, indicating that MusicBERT can perform well on song-level tasks.

\subsection{Method Analysis}
In this subsection, we analyze the effectiveness of each design in MusicBERT, including OctupleMIDI encoding, bar-level masking strategy, and the pre-training itself. We conduct experiments on MusicBERT\textsubscript{small} with a maximum sequence length of 250 due to the huge training cost of MusicBERT\textsubscript{base}. For simplicity, we treat melody completion and accompaniment suggestion as binary classification tasks: classifying matched and not matched melody or accompaniment pairs, instead of ranking a group of pairs by predicted match score. We use the accuracy percentage score as the metric for these two binary classification tasks.

\paragraph{Effectiveness of OctupleMIDI} We compare our proposed OctupleMIDI encoding with REMI~\cite{huang2020pop} and CP~\cite{hsiao2021compound} by training  MusicBERT\textsubscript{small} models with each encoding respectively and evaluate on downstream tasks. As shown in Table~\ref{tab:ablation-encoding}, for song-level tasks (i.e., genre and style classification), OctupleMIDI significantly outperforms REMI and CP based encoding, since the model can learn from a larger proportion of a music song with the compact OctupleMIDI encoding, given all encoding methods use the same length of sequence for pre-training. For phrase-level tasks (melody completion and accompaniment suggestion), the input sequence length is usually less than the truncate threshold. Thus, benefiting from the short representation, OctupleMIDI significantly reduces the computational complexity of the Transformer encoder, which is only 1/16 of that with REMI-like encoding and 1/4 of that with CP-like encoding. Moreover, according to Table~\ref{tab:ablation-encoding}, OctupleMIDI performs better than the other two encoding methods on phrase-level tasks.
\begin{table}[h]
\small
    \centering
    \begin{tabular}{lcccc}
        \toprule
        Encoding & Melody & Accom. & Genre & Style \\
        \midrule
        CP-like & 95.7 & 87.2 & 0.719 & 0.510 \\
        REMI-like & 92.0 & 86.5 & 0.689 & 0.487 \\
        \midrule
        OctupleMIDI & \textbf{96.7} & \textbf{87.9} & \textbf{0.730} & \textbf{0.534} \\
        \bottomrule
    \end{tabular}
    \caption{Results of different encoding methods. ``Accom.'' represents accompaniment suggestion task.}
    \label{tab:ablation-encoding}
\end{table}

\paragraph{Effectiveness of Bar-Level Masking} We compare our proposed bar-level masking strategy with two other strategies: 1) Octuple masking, as mentioned in Sec.~\ref{sec_mask_strategy} and Fig. \ref{fig:OctupleMIDI}; 2) Random masking, which randomly masks the elements in the octuple token similar to the masked language model in BERT. We use the same masking ratio for all these strategies. As shown in Table~\ref{tab:ablation-mask}, our proposed bar-level masking can effectively boost results on downstream tasks.

\begin{table}[h]
\small
    \centering
    \begin{tabular}{lcccc}
        \toprule
        Mask & Melody & Accom. & Genre & Style \\
        \midrule
        Random & 96.3 & 87.8 & 0.708 & 0.533 \\
        Octuple & 96.0 & 87.3 & 0.722 & 0.530 \\
        \midrule
        Bar & \textbf{96.7} & \textbf{87.9} & \textbf{0.730} & \textbf{0.534} \\
        \bottomrule
    \end{tabular}
    \caption{Results of different masking strategies.}
    \label{tab:ablation-mask}
\end{table}

\paragraph{Effectiveness of Pre-training}
To show the advantage of pre-training in MusicBERT, we compare the performance of MusicBERT\textsubscript{small} with and without pre-training. As shown in Table~\ref{tab:ablation-pretrain}, pre-training achieves much better scores on the four downstream tasks, demonstrating the critical role of pre-training on symbolic music understanding.

\begin{table}[ht]
\small
    \centering
    \begin{tabular}{lcccc}
        \toprule
        Model & Melody & Accom. & Genre & Style \\
        \midrule
        No pre-train & 92.4 & 76.9 & 0.662 & 0.395 \\
        \midrule
        MusicBERT & \textbf{96.7} & \textbf{87.9} & \textbf{0.730} & \textbf{0.534} \\
        \bottomrule
    \end{tabular}
    \caption{Results with and without pre-training.}
    \label{tab:ablation-pretrain}
\end{table}

\section{Conclusion}

In this paper, we developed MusicBERT, a large-scale pre-trained model for symbolic music understanding. Instead of simply adopting the pre-training methods from NLP to symbolic music, we handle the distinctive challenges in music pre-training with several careful designs in MusicBERT, including the efficient and universal OctupleMIDI encoding, the effective bar-level masking strategy, and the large-scale symbolic music corpus with more than 1 million music songs. MusicBERT achieves state-of-the-art performance on all of the four evaluated symbolic music understanding tasks, including melody completion, accompaniment suggestion, genre classification, and style classification. Method analyses also verify the effectiveness of each design in MusicBERT. For future work, we will apply MusicBERT on other music understanding tasks such as chord recognition and structure analysis to boost the performance.

\bibliographystyle{acl_natbib}
\bibliography{acl2021}

\end{document}